\documentstyle[din_a4,12pt]{article}
\begin{document}
\begin{center}
\Large
{\bf A Grand Canonical Ensemble Approach to the Thermodynamic Properties of the Nucleon 
in the Quark-Gluon Coupling Model } \\
\large
\vspace{1.0cm}
Hai Lin \\
\small{(April 2001)}
\\
 \vspace{0.5cm}
\normalsize
 $Department$ $of$ $ Physics,$ $ Peking $ $University,$ $ P.R.China,$ $ 100871$\\
{$ Email:hailin@mail.phy.pku.edu.cn$} \\
\vspace{0.5cm}
\large
{\bf Abstract} \\
\end{center}
\vspace{-0.3cm}
\normalsize
\small
In this paper, we put forward a way to study the nucleon's thermodynamic properties such as its temperature,
entropy and so on, without inputting any free parameters by human hand, even the nucleon's mass
and radius. First we use the Lagrangian density of the quark gluon coupling fields to deduce the Dirac Equation 
of the quarks confined in the gluon fields. By boundary conditions we solve the wave functions and energy eigenvalues 
of the quarks, and thus get energy-momentum tensor, nucleon mass, and density of states. Then we utilize a hybrid
grand canonical ensemble, to generate the temperature and chemical potentials of quarks, antiquarks of three 
flovars by the four conservation laws of the energy and the valence quark numbers, after which, all other thermodynamic 
properties are known. The only seemed free paremeter, the nucleon radius is finally determined by the grand 
potential minimal principle.
\vspace{1.0cm}
\normalsize

In our Quark Gluon Coupling Model, the nucleons are described as confining the quarks $(q)$ and gluons 
$(g)$ inside them, which are the quanta of the fields. The quarks interact through the exchange of gluons, 
and the gluon couples to the conserved quark-current through $g\overline{\psi }\gamma _{\mu }\psi A^{\mu }$,
where $g$ is the coupling constant.
The Lagrangian density for this model is 
\begin{eqnarray}
{\cal L}=\overline{\psi }\left[ \gamma ^{\mu }\left( i\partial _{\mu}-
gA_{\mu }\right) -m_{q}\right] \psi -\frac{1}{4}F_{\mu \nu }F^{\mu \nu}  
\end{eqnarray}
where $ F_{\mu \nu }=\partial _{\mu }A_{\nu }-\partial _{\nu }A_{\mu }. $

Lagrange's equations yield the the Dirac equation with the vector field[1,2]: 
\begin{equation}
\left[ \gamma ^{\mu }\left( i\partial _{\mu }-g A_{\mu }\right)- m_{q}\right] \psi =0, 
\end{equation}
and the equation of conserved quark current :
\begin{equation}
\partial _{\mu }{F^{\mu \nu }} =g \overline{\psi } \gamma ^{\nu }\psi.
\end{equation}

The energy-momentum tensor is: 
\begin{eqnarray}
T^{\mu \nu } &=&\frac{1}{4}F_{\lambda \sigma }F^{\lambda\sigma }
+i\overline{\psi }\gamma ^{\mu }\partial ^{\nu }\psi +\partial ^{\nu }A_{\lambda }F^{\lambda \mu }.
\end{eqnarray}

Lagrange's equations ensure that this tensor is conserved and satisfies $%
\partial _{\mu }T^{\mu \nu }=0$. It follows that the energy-momentum $P^{\nu}$ defined by 
\begin{equation}
P^{\nu }=\int d^{3}xT^{0\nu }
\end{equation}
which is a constant of motion.

We observe that at high quark density, the vector field operators can be replaced
 by their expectation values, which then serve as classical, condensed fields in which the quarks move, 
\begin{eqnarray}
A_{\mu  }\rightarrow \delta _{\mu 0}A_{0}.
\end{eqnarray}
For a static, uniform system, the quantities $A_{0}$ are 
constants independent of $x_{\mu }$. In the Mean Field Theory, the Lagrangian density is 
\begin{eqnarray}
{\cal L} &=&\overline{\psi }\left[ i\gamma ^{\mu }\partial
_{\mu }-g\gamma ^{0}A_{0}-m_{q}\right] \psi 
\end{eqnarray}

Hence, the Dirac equation is linear, 
\begin{equation}
\left[ i\gamma ^{\mu }\partial _{\mu }-g\gamma ^{0}A_{0}-m_{q} \right] \psi =0.
\end{equation}
We seek normal-mode solutions of the form $\psi
(x^{\mu })=\psi (\vec{r})e^{-iEt}$. This leads to 
\begin{eqnarray}
[-i\vec{\alpha}.\nabla +gA_{0}+\beta m_{q}]\psi (\vec{r}) =E\psi (\vec{r})
\end{eqnarray}

Consider the case $k=-1$, which is the $S_{1/2}$ level.
The normalized quark wave function for a sphere of radius $R$ is[3]:
\begin{equation}
{\psi}_q({\vec{r}},t) = {\cal{N}} e^{-i {\epsilon}_q t}
\times \left( \begin{array}{c} { \sqrt{\frac{{\epsilon}_q-g{A}_0+{m}_q}{{\epsilon}_q}}
  j_0( \sqrt{{({\epsilon}_q-g{A}_0)}_{}^{2}-{m}_{q}^{2}}r)} \\
i  { \sqrt{\frac{{\epsilon}_q-g{A}_0-{m}_q}{{\epsilon}_q}} j_1( \sqrt{{({\epsilon}_q-g{A}_0)}_{}^{2}-{m}_{q}^{2}}r))
  }\end{array} \right)
{{{\chi}_q}\over{\sqrt{4 \pi}}} ,
\end{equation}
where $r=|{\vec{r}}|$, ${\chi}_q$ is the quark spinor and ${\cal{N}}$ is the normalization constant[4]. 
${\epsilon}_q$ is the quark energy eigenvalue.
The density of quarks is readily calculated as 
\[
J^{0}=\overline{\psi }\gamma ^{0}\psi  \left[ j_{0}^{2}\left(\sqrt{{({\epsilon}_q-g{A}_0)}_{}^{2}-{m}_{q}^{2}}r\right)
 +\frac{{\epsilon}_q-g{A}_0-{m}_q} {{\epsilon}_q-g{A}_0+{m}_q}\ j_{1}^{2}
\left( \sqrt{{({\epsilon}_q-g{A}_0)}_{}^{2}-{m}_{q}^{2}}r%
\right) \right] \theta _{V}, 
\]
where 
\[
\theta _{V}=\left\{ 
\begin{array}{l}
1\qquad r\leq R \\ 
0\qquad r>R.
\end{array}
\right. 
\]

Thus, the density certainly does not vanish at $r=R$. Clearly, although the
lower component is suppressed for small $r$, it does make a sizeable
contribution near the surface of the nucleon. However, it is not the density, but $\overline{\psi }\psi $ should vanish
at the boundary in the relativistic theory[5], 
\[
\left. \overline{\psi }\psi \right| _{r=R}=\frac{{\epsilon}_q-g{A}_0+{m}_q}{{\epsilon}_q}\ j_{0}^{2}
(\sqrt{{({\epsilon}_q-g{A}_0)}_{}^{2}-{m}_{q}^{2}}r )-%
\frac{{\epsilon}_q-g{A}_0-{m}_q}{{\epsilon}_q}\ j_{1}^{2}(\sqrt{{({\epsilon}_q-g{A}_0)}_{}^{2}-{m}_{q}^{2}}r )=0. 
\] 
 This yields the energy states
\begin{equation}
{\epsilon}_{q,n}^{}=g{A}_0+\sqrt{{m}_{q}^{2}+{(\frac{n\pi}{R})}_{}^{2}} ,
\end{equation} 
 where $n=1,2,...$  and the density of states is
\begin{equation}
{\rho}_q(\epsilon)=\frac{R(\epsilon-g{A}_0)} {\pi \sqrt{{(\epsilon-g{A}_0)}_{}^{2}-{m}_{q}^{2}}}.
\end{equation}

Then we consider the energy-momentum tensor for the following type, 
\[
T_{V}^{\mu \nu }=T^{\mu \nu }\theta _{V}, 
\]
and $T^{\mu \nu }$ is the familiar energy-momentum tensor for a free Dirac
field $ T^{\mu \nu }=i\overline{\psi }\gamma ^{\mu }\partial^{\nu }\psi$. 

The condition for overall energy and momentum conservation is that the
divergence of the energy-momentum tensor should vanish, and this is
certainly true for $T^{\mu \nu }$, as is easily proven from the free Dirac
equation $ \partial _{\mu }T^{\mu \nu }=0.$ 

However, the fact that the quarks move only inside the restricted
region of space $V$ leads to problems. Indeed,$ \partial _{\mu }\theta _{V}=n_{\mu }\Delta _{s},$ 
where $\Delta _{s}$ is a surface delta function $\Delta _{s}=-n.\partial (\theta _{V})$ 
and $ {n}_{}^{\mu}=(0, \hat r ) $.  In the static spherical case we find that $\Delta _{s}$ is simply $\delta
(r-R)$. Putting all these together we obtain $ 
\partial _{\mu }T_{V}^{\mu \nu }=i\overline{\psi }\gamma .n%
{\partial }^{\nu }\psi \Delta _{s},$
and using the linear boudary condition, 
\[
\partial _{\mu }T_{V}^{\mu \nu }=-\frac{1}{2}\left. \partial ^{\nu }(%
\overline{\psi }\psi )\right| _{s}\Delta _{s}=-Pn^{\nu }\Delta _{s}, 
\]
where $P$ is the pressure exerted on the sphere's surface by the contained Dirac gas 
\[
P=-\frac{1}{2}\left. n.\partial ^{\nu }(\overline{\psi }\psi )\right| _{s}. 
\]
To keep the energy-momentum conservation, we add an energy density term $B\theta _{V}$
to the Lagrangian density. Then (since $T^{\mu \nu }$ involves ${\cal L}%
g^{\mu \nu }$) the new energy-momentum tensor $T_{2}^{\mu \nu }$
has the form 
\[
T_{2}^{\mu \nu }=(T^{\mu \nu }+Bg^{\mu \nu })\theta _{V}. 
\]
Therefore, the divergence of the energy-momentum tensor is 
\[
\partial _{\mu }T_{2}^{\mu \nu }=(-P+B)n^{\nu }\Delta _{s}, 
\]
which will vanish if 
\begin{equation}
B=P=-\frac{1}{2}\left. n.\partial ^{\nu }(\overline{\psi }\psi )\right|
_{s}. 
\end{equation} 

Thus the nucleon's mass is given by
\begin{equation}
M =\int d{x}_{}^{3}\theta _{V}T^{00}+{{4}\over{3}} \pi B R^3 .
\label{BAGENERGY}
\end{equation}

Then we study the nucleon's thermodynamic properties, considering it as a many-body system at a given 
temperature ($T$). This is a hybrid system within radius $R$ of 
interacting massive quarks, antiquarks of 3 flavors  $(u,\bar{u},d,\bar{d},s,\bar{s})$ and
 massless gluons $(g)$ , all of which can exchange the energy and the 
particle numbers[6].  In a statistical approach, the grand canonical partition function is given by
\begin{equation}
Z_G \ = \ Z_{\rm {vac}}\ Z_{u}\ Z_{\bar{u}}\ Z_{d}\ Z_{\bar{d}} \ Z_{s} \ Z_{\bar{s}} \ Z_{g}, 
\end{equation}

\noindent where $ Z_{\rm {vac}} $ takes care of the temperature $ T \rightarrow 0 $ limit, and we consider 
this as
\begin{equation}
-T \ln Z_{\rm {vac}} \ = \ {{4}\over{3}} \pi B R^3.
 \   \label{surf2}
\end{equation}
$ Z_{u} $, $ \ Z_{d} $, $Z_{ s}$, $\ Z_{\bar{u}}$, $\ Z_{\bar{d}}$ and $\ Z_{\bar{s}}$ refer
 to the partition function for the u,d,s quark and $\bar{u}$, $\bar{d}$, $\bar{s}$ antiquarks, while
$Z_{g} $ refers to the gluonic part. 
Their chemical potentials obey that $ \mu(u)=-\mu(\bar{u}) $, $ \mu(d)=-\mu(\bar{d}) $, $ \mu(s)=-\mu(\bar{s}) $
and $ \mu(g)=0 $, among which there are only 3 free parameters. 

The total nubmer, energy and grand partition function of each quark system are:
\begin{equation}
N_q  \ = \ {\sum_i} \ {1 \Big{/}{\Big(} \ {e}_{}^{({\epsilon}_{q,i}- \mu_q )/T}\ + \ 1{\Big)}}= 6
\int_{m_q+gA_0}^{\infty} \ {1 \Big{/}{\Big(} \ {e}_{}^{(\epsilon- \mu_q )/T}\ + \ 1{\Big)}} \rho_q(\epsilon) d\epsilon, 
\end{equation}
\begin{equation}
E_q  \ = \ {\sum_i} \ {\epsilon_i \Big{/}{\Big(} \ {e}_{}^{({\epsilon}_{q,i}- \mu_q )/T}\ + \ 1{\Big)}}=6
\int_{m_q+gA_0}^{\infty} {1 \Big{/}{\Big(} \ {e}_{}^{(\epsilon- \mu_q )/T}\ + \ 1 {\Big)}} \rho_q(\epsilon)
\epsilon  d\epsilon, 
\end{equation}

\begin{equation}
 \ln Z_q \ = \  {\sum_i} \ \ln \ {\Big (} \ 1  \ + \ {e}_{}^{- ({\epsilon_{q,i} - \mu_q)/T}} {\Big)}=6
\int_{m_q+gA_0}^{\infty}  \ \ln \ {\Big (} \ 1  \ + \ {e}_{}^{- ({\epsilon - \mu_q)/T}} {\Big)}\rho_q(\epsilon) d\epsilon. 
\end{equation}

And the total energy and grand partition function of the gluon system are $E_g=
{{64{\pi}_{}^{5}}\over{15}}T_{}^{4}V$ and  $ \ln Z_g= {{64{\pi}_{}^{5}}\over{45}}T_{}^{3}V$. 
We obtain the four parameters  $ \mu(u) $, $ \mu(d) $, $ \mu(s) $ and $T$
 by the following four conservation laws[7]:
\begin {eqnarray}
N_u-N_{\bar{u}}=n(u),  \nonumber 
\end {eqnarray}
\begin {eqnarray}
N_d-N_{\bar{d}}=n(d), \nonumber 
\end {eqnarray}
\begin {eqnarray}
N_s-N_{\bar{s}}=0,    \nonumber 
\end {eqnarray}
and
\begin {eqnarray}
E_u+E_{\bar{u}}+E_d+E_{\bar{d}}+E_s+E_{\bar{s}}+E_g=M-{{4}\over{3}}\pi B R^3.\nonumber 
\end {eqnarray}
Therefore the grand potential of the hybrid system is given by
\begin{equation}
\Omega \  =-T \ln Z_G
\end{equation}
Thus we get the entropy of the nucleon as:
\begin{equation}
S=-(\partial{\Omega}/\partial T){|}_{\mu(u),\mu(d),\mu(u),V}^{}
\end{equation}
At last, the only unsettled parameter, the nucleon's radius $R$, that we input into the model at the 
beginning could be determined by the grand potential minimal principle:
\begin{equation}
(\partial{\Omega}/\partial R){|}_{\mu(u),\mu(d),\mu(u),T}^{}=0
\end{equation}

\end{document}